\documentstyle[aps]{revtex}
\begin{document}
\title{Hall resistance in the hopping regime, a ``Hall Insulator"?}
\draft
\author{O. Entin-Wohlman}
\address{School of Physics and Astronomy,
Raymond and Beverly Sackler Faculty of Exact Sciences\\
Tel Aviv University, Tel Aviv 69978, Israel}
\author {A. G.  Aronov\cite{AGA}, Y. Levinson, and Y. Imry}
\address{Condensed Matter Department, The Weizmann Institute of Science\\
Rehovot 76100, Israel}
\date{\today}
\maketitle
\begin{abstract}
The Hall conductivity and  resistivity of strongly localized electrons
 at
low temperatures and at small
 magnetic fields are obtained.
It is found that the results depend on whether the conductivity or the
resistivity tensors  are averaged to obtain the macroscopic Hall resistivity.
In the second case the Hall
resistivity always {\it diverges}
exponentially as the temperature tends to zero. But when the Hall
resistivity is derived from the
averaged conductivity, the resulting temperature dependence is
sensitive to the disorder
configuration. Then the Hall resistivity may approach a constant
 value as
$T\rightarrow 0$. This is the Hall
insulating behavior.  It is argued that
for strictly dc conditions, the transport quantity that should
 be averaged is the resistivity.
\end{abstract}
\pacs{72.20.My, 72.20.-i, 71.50+t}
Recently it has been stated
\cite{zhang,viehweger,imry}  that the
zero-temperature Hall resistivity
$\rho_{xy}(\omega )$ of noninteracting electrons in the insulating
 regime remains
finite as the frequency $\omega\rightarrow 0$. This result was derived
 from the
Kubo formula for the frequency-dependent conductivity. It was found
 that at
zero temperature the disorder-averaged
$\sigma_{xy}(\omega )$ of the Anderson insulator vanishes at low
 frequencies
proportionally to $\omega ^{2}$. Since to leading order in $\omega $ the
longitudinal conductivity
$\sigma_{xx}\sim
i\omega\varepsilon_{0}$, where $\varepsilon_{0}$ is the dielectric constant,
this yielded that the Hall
resistivity $\rho_{xy}\sim\sigma_{xy}/(\sigma_{xx}^{2}+\sigma_{xy}^{2})$
approaches a
constant in the small-frequency, zero-temperature limit.

Obviously one would like to examine
the Hall resistance at finite temperatures. Here we address this
 question
for strongly localized electrons.
We consider the problem using the Holstein model
\cite{holstein} for the Hall effect in a system with localized states.
The conductivity tensor for a triangle of three sites is obtained first.
To get the macroscopic Hall resistivity one may attempt to apply two
 averaging procedures. We find that the result strongly depends
on whether the conductivity tensor is averaged before inverting it
or {\it vice versa}. In the first case a Hall insulator behavior is possible,
 while in the second the macroscopic Hall resistivity
increases with decreasing temperature. The latter is similar to the
finding of Friedman and Pollak
\cite{friedman}. The two behaviors are argued
to be related with ``ac" and ``dc" conditions.

The previous \cite{zhang,viehweger,imry} discussions of the zero-temperature
Hall resistivity were all based
on the ensemble averaged Kubo formula, which yields that
${\bar\sigma}_{xx}={\bar\sigma}_{yy}$ and that
${\bar\sigma}_{xy}$ vanishes proportionally to the magnetic
field (the bar above the quantity indicates ensemble averaging).
 However, before averaging,
$\sigma_{xy}$ includes a field-independent term. We show that in
 the strongly
localized regime this term is comparable in magnitude to
$\sigma_{xx}$ and $\sigma_{yy}$. This leads to delicate  cancellations
 when the
local (unaveraged) conductivity tensor is inverted to obtain the
resistivity, and
consequently to the very different temperature dependences described
 above.

It is convenient to investigate the transport properties of electrons
 in
the hopping regime by constructing the
rate equations for the electron distributions, utilizing the electronic
transition probabilities between
localized states. One has
\begin{equation}
\frac{dn_{i}}{dt}=\sum_{j}(P_{ji}-P_{ij}),\label{1}
\end{equation}
in which $n_{i}$ is the electronic population of site $i$ (the term
 `site'
is used for a localized state)
and $P_{ij}$ is the rate of the population
decay by phonon-assisted hopping into site $j$. A delicate point
 is the
dependence of the rate
on the magnetic field, $H$. As was
shown by Holstein \cite{holstein}, this is due to interference of
 the ``direct"
amplitude to hop from $i$ to $j$,  with the indirect amplitude via
 a third
site $\ell$,
$i\rightarrow \ell\rightarrow j$. The magnetic field dependence of $P_{ij}$
then necessitates the
consideration of at least three sites. To write this rate explicitly, we
employ the Holstein model \cite{polaron}
for the electron-phonon interaction, in which the ion displacements are
coupled to the local (site) density of
the electrons, and denote by $\epsilon_{i}$ the single-particle energies of
the localized
states, which are assumed to be randomly distributed, and by $J_{ij}$ the
overlap of two wave functions localized
at sites $i$ and $j$. The strong localization regime is characterized by
\cite{holstein,ambegaokar} $\mid J_{ij}\mid\ll
\mid\epsilon_{ij}\mid ,\epsilon_{ij}=\epsilon_{i}-\epsilon_{j}$. One finds
\begin{equation}
P_{12}=P_{12}^{dir}+P_{12}^{indir}.\label{2}
\end{equation}
Here $P_{12}^{dir}$ arises from the direct amplitude alone and is
independent of the magnetic field,
\begin{eqnarray}
P_{12}^{dir}&=&n_{1}(1-n_{2})\mid J_{12}\mid^{2}
\int_{-\infty}^{\infty}dt\ e^{i\epsilon_{12}t}\ e^{2(g(t)-g(0))},\nonumber\\
g(t)&=&\sum_{q}\frac{\mid v_{q}\mid^{2}}{\omega_{q}^{2}}[(1+N_{q})\
e^{-i\omega_{q}t}+N_{q} \ e^{i\omega_{q}t}],
\label{3}
\end{eqnarray}
where $\omega_{q}$ is the phonon frequency, $v_{q}$ is the interaction
matrix element and
$N_{q}=1/(e^{\beta\omega_{q}}-1)$ ($\hbar =1$).
$P_{12}^{indir}$ comes from the interference between the direct
($1\rightarrow 2$) and the indirect
($1\rightarrow 3\rightarrow 2$) amplitudes
\begin{eqnarray}
&\ &P_{12}^{indir}=n_{1}(1-n_{2})\mid J_{12}J_{23}J_{31}\mid
 2\Im\ e^{i\phi_{132}}\nonumber\\
&\ &\int_{0}^{\infty}dt_{2}\int_{-\infty}^{\infty}
dt_{1}\ e^{g(t_{1})+g(t_{2})+g(t_{1}+t_{2})-3g(0)}\nonumber\\
&\ &\Bigl[(1-n_{3})e^{i\epsilon_{12}t_{1}+i\epsilon_{13}t_{2}}
-n_{3}\ e^{i\epsilon_{12}t_{1}+i\epsilon_{32}t_{2}}\Bigr],\label{4}
\end{eqnarray}
and contains the occupation of site $3$, $n_{3}$, and the magnetic phase,
$\phi_{132}$,
\begin{eqnarray}
\phi_{132}&=&\frac{e}{c}\vec{H}\cdot\vec{S},\nonumber\\
\vec{S}&=&(\vec{R}_{1}\times\vec{R}_{3}+\vec{R}_{3}\times
\vec{R}_{2}+\vec{R}_{2}\times\vec{R}_{1})/2.\label{5}
\end{eqnarray}
Here $\vec{R}_{i}$ is the radius vector of
site $i$ and $\vec{S}$ is the vectorial area of the triangle.
The field-dependent part of $P_{12}$ includes a term
even in the field (proportional to $\cos\phi_{132}$) and a term odd in it
(proportional to $\sin\phi_{132}$). The first gives a correction to the
direct rate and will be discarded henceforth. Obviously it is the term
odd in the field which is capable of producing the Hall resistance.

We now apply the rate equation (\ref{1}) to a group of three sites, $1$,
$2$ and $3$,
to obtain the current driven by an external ac electric field $\vec{E}$ of
small frequency $\omega $.
In the presence of an electric field the
occupations $n_{i}$ will be modified in a way that can be expressed by
changes $\delta\mu_{i}$ in the local
chemical potentials \cite{ambegaokar}
\begin{equation}
n_{i}=n_{i}^{0}-\beta n_{i}^{0}(1-n_{i}^{0})\delta\mu_{i},\label{6}
\end{equation}
where $n_{i}^{0}$ is the Fermi distribution. Also, the rate $P_{ij}$
 which
depends on $\epsilon_{i}-\epsilon_{j}$
is changed to depend on
$\epsilon_{i}-\epsilon_{j}+e\vec{E}\cdot(\vec{R}_{i}-\vec{R}_{j})$. This
way one
obtains the response of the system to the local electrochemical potential
differences \cite{ambegaokar}
\begin{equation}
\zeta_{ij}=\frac{\delta\mu_{i}-\delta\mu_{j}}{e}-\vec{E}\cdot\vec{R}_{ij},
\
\vec{R}_{ij}=\vec{R}_{i}-\vec{R}_{j}.\label{7}
\end{equation}
Explicit calculations of Eqs. (\ref{3}) and (\ref{4}) yield
\begin{eqnarray}
e\frac{dn_{1}}{dt}&=&e(P_{21}-P_{12}+P_{31}-P_{13})\nonumber\\
&=&G_{12}\zeta_{12}+G_{31}\zeta_{13}+\sin\phi\Gamma\zeta_{23},\label{8}
\end{eqnarray}
with analogous expressions for the other two sites, where $\phi $ is the
magnetic flux enclosed in the
triangle. These equations determine the electrochemical potential differences,
$\zeta_{ij}$. However, in practice one does not have to solve for the
$\zeta_{ij}'s$ since Eq. (\ref{8}) gives
the current \cite{ambegaokar} (see Eq. (\ref{13}) below.)
In (\ref{8}), $G_{ij}$ is the conductance of the bond $ij$, arising from
the direct rate,
\begin{eqnarray}
&\ &G_{ij}=G_{ji}=\nonumber\\
&\ &e^{2}\beta n_{i}^{0}(1-n_{j}^{0})\mid J_{ij}\mid^{2}
\int_{-\infty}^{\infty}dt\
e^{i\epsilon_{ij}t}\ e^{2(g(t)-g(0))}.\label{9}
\end{eqnarray}
At low temperatures \cite{ambegaokar} and for weak electron-phonon coupling,
the bond conductance becomes
\begin{equation}
G_{ij}=e^{2}\beta\mid J_{ij}\mid^{2}\frac{v^{2}{\cal
N}(\mid\epsilon_{ij}\mid )}{\mid\epsilon_{ij}\mid^{2}}
e^{-\frac{\beta
}{2}(\mid\epsilon_{i}\mid+\mid\epsilon_{j}\mid+\mid\epsilon_{i}-\epsilon_{j}
\mid )},\label{10}
\end{equation}
where ${\cal N}$ is the phonon density of states. (Energies are measured from
the Fermi energy.)
The interference process leads to the phase-dependent term, with
\begin{eqnarray}
&\ &\Gamma =e^{2}\beta n_{1}^{0}(1-n_{2}^{0})
\mid J_{12}J_{23}J_{31}\mid \nonumber\\
&\ &\int_{-\infty}^{\infty}dt_{1}\int_{-\infty}^{\infty}dt_{2}\
e^{g(t_{1})+g(t_{2})+g(t_{1}+t_{2})-3g(0)}\nonumber\\
&\ &e^{i\epsilon_{12}t_{1}}\Bigl((1-n_{3}^{0})e^{i
\epsilon_{13}t_{2}}+n_{3}^{0}e^{i\epsilon_{32}t_{2}}\Bigr).\label{11}
\end{eqnarray}
Expanding this expression for weak electron-phonon interaction and using
(\ref{10})
we find that $\Gamma$ can be
written in the form
\begin{eqnarray}
&\ &\Gamma =\frac{\mid
J_{12}J_{23}J_{31}\mid}{4e^{2}\beta}\Bigl(\frac{G_{31}G_{12}}{\mid
J_{31}J_{12}\mid^{2}n_{1}^{0}(1-n_{1}^{0})}
\nonumber\\
&+&\frac{G_{12}G_{23}}{\mid J_{12}J_{23}\mid^{2}n_{2}^{0}(1-n_{2}^{0})}+
\frac{G_{23}G_{31}}{\mid
J_{23}J_{31}\mid^{2}n_{3}^{0}(1-n_{3}^{0})}\Bigr).\label{12}
\end{eqnarray}
This expresses the fact that the indirect rate involves {\it two}
scattering events by the phonons \cite{holstein,we}. The temperature dependence
of $\Gamma $, at low
temperatures, can be obtained from (\ref{10}), using
$n_{i}^{0}(1-n_{i}^{0})\sim\exp[-\beta\mid\epsilon_{i}\mid]$.

The current density is given by
\begin{equation}
\vec{j}=e\sum_{i}\frac{dn_{i}}{dt}\vec{R}_{i}.\label{13}
\end{equation}
If one now introduces the effective field $\vec{E}_{eff}$ which produces
the electrochemical potential,
\begin{eqnarray}
\zeta_{ij}&=&-\vec{E}_{eff}\cdot\vec{R}_{ij}, \nonumber\\
\vec{E}_{eff}&=&\frac{1}{2 S_{z}}
[\zeta_{23}\vec{R}_{1}+\zeta_{31}\vec{R}_{2}+\zeta_{12}\vec{R}_{3}]\times\hat{z},
\label{14}
\end{eqnarray}
the current density becomes
\begin{eqnarray}
\vec{j}&=&\tensor{\sigma}\vec{E}_{eff}\nonumber\\
\tensor{\sigma}&=&\vec{R}_{12}\vec{R}_{12}G_{12}+
\vec{R}_{23}\vec{R}_{23}G_{23}+\vec{R}_{31}\vec{R}_{31}G_{31}\nonumber\\
&+&\Gamma\sin\phi(\vec{R}_{23}\vec{R}_{1}+\vec{R}_{31}\vec{R}_{2}+\vec{R}_{12}\vec{R}_{3}).\label{15}
\end{eqnarray}
A remarkable observation is that the part of $\vec{j}$ which is
proportional to the magnetic field is
perpendicular to the effective electric field. In deriving the expression for
$\tensor{\sigma}$ it was assumed
that the triangle lies in the $x-y$ plane, perpendicular to $\hat{z}$. Note
that
$\vec{E}_{eff}$, Eq. (\ref{14}),
is invariant to the choice of the coordinate origin.

The current response to the electrochemical potential difference
is our central result. From (\ref{15}) one finds for the conductivity tensor
\begin{equation}
\sigma_{xx}=(R_{12}^{x})^{2}G_{12}+(R_{23}^{x})^{2}G_{23}+(R_{31}^{x})^{2}G_{31},\label{16}
\end{equation}
($\sigma_{yy}$ is given upon replacing $R_{ij}^{x}$ by $R_{ij}^{y}$), and
\begin{eqnarray}
\sigma_{\stackrel{xy}{yx}}
&=&R_{12}^{x}R_{12}^{y}G_{12}+R_{23}^{x}R_{23}^{y}G_{23}+R_{31}^{x}R_{31}^{y}G_{31}\nonumber\\
&\pm & 2 S_{z}\Gamma\sin\phi.\label{17}
\end{eqnarray}

It is straightforward to check that these results follow very simply also
from the
Kubo formulation, with the understanding that it yields the response to the
{\it effective field}, $\vec{E}_{eff}$:
\begin{eqnarray}
\sigma_{ij}(\omega)&=&\frac {ie^2\omega}{ Vol} \sum_{m,n}
\Big[ \frac{<m|x_i|n><n|x_j|m>}{\omega-\epsilon_{nm}+i\eta}\nonumber\\
&-&\frac{<m|x_j|n><n|x_i|m>}{\omega+\epsilon_{nm}+i\eta}
\Big] (f_m-f_n),\label{18}
\end{eqnarray}
where  $\eta$ is a positive infinitesimal, $|m>$ and $|n>$ are eigenstates
and $f_n$, $f_m$ their populations.
We generalize the derivation of \cite{imry} to include electron-phonon coupling
and consider  the $\omega>0$,
$\omega \rightarrow 0$ limit.  For example, to get the leading, (1,2)-type
terms, in (16)  and in the first line of
(17), the two  relevant states are: $|1>$: electron in site 1 with
 a phonon bath
in equilibrium  and $|2,q>$:
electron in site 2, with the same minus one phonon
in state $q$, where $\omega_q \cong \epsilon_{21}$.  Thus, the
approximate eigenstates are:
\begin{equation}
|m> =|1> + \sum_q \frac{J_{12}\nu_q}{\epsilon_{12}+\omega_q} |2,q> ,\label{19}
\end{equation}
\begin{equation}
|n> = |2,q> + \frac {J_{21}\nu^*_q}{\epsilon_{21}-\omega_q} |1>,
\label{20}
\end{equation}
where $\nu_q = v_q \sqrt{N_q}/\omega_q$, and
\begin{equation}
<n|x|m> \cong \frac{J_{12}|\nu_q|^2R^x_{12}}{\omega}.\label{21}
\end{equation}
Using Eq. (\ref{18}), this produces the first term in (\ref{16}):
\begin{eqnarray}
&\ &\frac{1}{Vol}(R^x_{12})^2 e^2\beta J^2_{12}\sum_q
\frac{|v_q|^2}{\omega_q^2}
N_{q}\pi\delta(\omega-\epsilon_{21}-\omega_q) =\nonumber \\
&\pi &\beta e^2 J^2_{12}(R^x_{12})^2
\frac{|v_q|^2}{\omega_q^2}N_{q}{\cal
N}(\epsilon_{21})\mid_{\omega_{q}=\epsilon_{21}}.\label{22}
\end{eqnarray}
Using the expression
similar to (\ref{20}) for the matrix elements of $y$, yields the first
 term in (\ref{17}). The (2,3) and (1,3) terms in Eqs. (\ref{16}) and
(\ref{17})
are similarly obtained.
To get
the $\Gamma$ term in (\ref{17}), one has to mix in Eq. (\ref{19})
also the state $|3,q,q'>$
(i.e. $|3>$ minus the phonons $q$ and $q')$ in two ways:

(a)  A straightforward correction $\sum_{q'}
[J_{13}\nu_q\nu_{q'}/(\epsilon_{13}+\omega_{q}+\omega_{q'})]|3,q,q'>$,
which is first order in $J$ but second order in the electron-phonon
interaction.
It will be dominated by the ``resonant" contribution ($q'$ such that
$\omega_q+\omega_{q'} =
\epsilon_{31}$) and by the nonresonant contribution given by
$q' = q"$ such that $\omega_{q"} = \epsilon_{32}$.

(b) The Holstein contribution,the mixing of $\mid 2,q>$ via the intermediate
state $\mid 3,q,q'>:$
$\sum_{q'}[J_{13}J_{32}\nu_{q}\mid \nu_{q'}\mid^{2}/
(\epsilon_{13}+\omega_{q}+\omega_{q'})(\epsilon_{12}+\omega_{q})]\mid 2,q>.$
Here, as found by Holstein, the resonant contribution with
$\epsilon_{31} =
\omega_{q}+\omega_{q'}$ will yield the needed phase to have  a term odd
 in the magnetic field (the
$\Gamma$ term) in (\ref{17}).
Putting all the above together we find within the required accuracy
\begin{eqnarray}
\langle n|x|m\rangle &\cong &
\frac{J_{12}\nu_{q}}{\omega}R^{x}_{12}-2i\pi\delta(\epsilon_{31}-\omega_{q}-\omega_{q'})\nonumber\\
&\ &J_{13}J_{32}\nu_{q}\mid\nu_{q'}\mid^{2}
\Big[R^{x}_{3}-\frac{R^{x}_{1}+R^{x}_{2}}{2}\Big].\label{23}
\end{eqnarray}
Using this in Eq. (\ref{18}) produces the additional $\Gamma$ term in Eq.
(\ref{17}).

The result of the above calculations is the conductivity tensor
$\tensor{\sigma}$
of a single triangle. We would like to have
macroscopic quantities like the Hall resistivity
$\rho_{xy}$. For that purpose
one might average $\tensor{\sigma}$ over the orientations and
sizes of the triangles and also over the on-site energies.
{\it A priori} two averaging procedures exist. One can average
$\tensor{\sigma}$,
resulting in $\bar \sigma$, and then calculate
 $\rho_{xy} \equiv (\bar \sigma^{-1})_{xy}.$
One can also average the resistivity tensor of a triangle
$\tensor{\sigma}^{-1}$
resulting in $\overline {\sigma^{-1}}$ and then calculate
 $\rho_{xy} \equiv ( \overline {\sigma^{-1}})_{xy}$.
As we will see these two procedures lead to qualitatively different results.

We note that
$\sigma_{xy}$ includes a term independent of the magnetic field,
 which is of the
same order of magnitude as
$\sigma_{xx}$. Were we to average
$\tensor{\sigma}$ over directions {\it before } inverting it, then this
term would
have disappeared. However, if $\tensor{\rho}$ is to be averaged, then this term
remains, and leads to delicate cancellations in the denominator of
$\tensor{\rho}$.
This, in turn, is the cause of the two different temperature dependences
 of the Hall resistivity.

 We first average the conductivity tensor over directions and then invert it.
In
that case,
\begin{equation}
\rho_{xy}^{(1)}=\frac{2S_{z}\Gamma\sin\phi}{(R_{12}^{x})^{4}G_{12}^{2}},\label{24}
\end{equation}
where we have assumed for simplicity that $G_{12}$ is the largest conductance.
To obtain the temperature dependence we consider the situation in which the
magnetic field-free hopping
conduction takes place along the bond $12$ and site $3$ supplies the
interference path. Thus we imagine
$\epsilon_{1}$ and $\epsilon_{2}$ to be below and above the Fermi level,
but close to it, while $\epsilon_{3}$ is
away from the Fermi energy. Then [{\it cf.} Eqs. (\ref{10}) and (\ref{12})]
\begin{equation}
\rho_{xy}^{(1)}\sim\frac{\Gamma}{G_{12}^{2}}\sim\exp[-\beta(
\epsilon_{3}-2\epsilon_{2}+\epsilon_{1
})],\label{25}
\end{equation}
and is sensitive to the averaging procedure over the on-site energy
distribution. One may
imagine that the energy $\epsilon_{2}$
is mostly in-between the energies $\epsilon_{1}$ and $\epsilon_{3}$, in
which case the log-average of
$\rho_{xy}^{(1)}$ will lead to a constant value for the Hall resistivity at
very low temperatures - i.e. a
``Hall-insulating behavior".

We next consider the transverse resistivity obtained by inverting the full
conductivity tensor and then
averaging over directions to restore rotational invariance. The result is
\begin{equation}
\rho_{xy}^{(2)}=\frac{1}{2S_{z}}\frac{\Gamma\sin\phi}{G_{12}G_{23}+G_{23}G_{
31}+G_{31}G_{12}}.\label{26}
\end{equation}
Due to the cancellations occuring when $\tensor{\sigma}$ of Eqs. (\ref{16})
and (\ref{17}) is inverted, the
denominator here includes $G_{12}G_{23}$, {\it etc.}, but not the (larger)
term $G_{12}^{2}$, as in (\ref{24}).
Consequently, $\rho_{xy}^{(2)}$
increases exponentially as the temperature tends to zero, independent of the
specific configuration of the single-particle energies. This is because of the
factors $n_{i}^{0}(1-n_{i}^{0})$ in Eq.
(\ref{12}). Consider for example
the energy configuration specified above.  In that case the leading
term in $\Gamma $ is of
 order
$\exp[-\beta (\epsilon_{3}-\epsilon_{1})]$ while $G_{12}G_{23}$
($G_{12}\sim\exp[-\beta(\epsilon_{2}-\epsilon_{1})]$,
$G_{23}\sim\exp[-\beta\epsilon_{3}]$) dominates the denominator in Eq.
(\ref{26}), leading to $\rho_{xy}^{(2)}\sim\exp[\beta\epsilon_{2}]$.

Both $\rho_{xy}^{(1)}$ and $\rho_{xy}^{(2)}$ are {\it independent} of
the strength of the coupling to the
phonons. This is in analogy with the ``classical" (Boltzmann equation)
result for the Hall coefficient, which
does not contain the mean free path.

The physically
correct way of averaging may depend on whether the experiment is a dc one or
an ac one. We consider a measurement to be an ac one if the frequency is
 such that the electron
 cannot traverse the sample from one current contact to the other during
one field period. In this case the current contacts are irrelevant,
the current is inside the sample, and the macroscopic current
density is obtained by summing the contributions from all triangles within
a unit volume. This summation is equivalent to the averaging of the
conductivity tensor of a single triangle.

We define a  measurement as a dc one when
the electron traverses the sample during a time
short compared to the period of the field. The
 current is therefore flowing from one current contact to the other
  through a percolation chain of bonds.
 This means that the direction of the current in each
bond is defined. To find the elementary Hall voltage created at this
bond we have to invert the conductivity tensor of the triangle
attached to this bond and express this elementary Hall voltage
in terms of the resistivity tensor of a single triangle.
The total Hall voltage is obtained in this case by
summing over the bonds along the
percolation chain. This summation is equivalent to the averaging of the
resistivity tensor of a single triangle.

To summarize:  Two independent derivations of $\tensor\sigma$ at
 zero frequency
but finite temperatures, were given for the Holstein model.
Here the Hall conductivity too has a finite dc value when real,
phonon-mediated, transitions are
allowed.  Surprisingly enough, we find that the answer depends on
 which
transport quantity is averaged over
directions. It was shown
that the ensemble averaging needed to get the macroscopic Hall resistivity
is subtle and the result depends on whether
$\tensor\sigma$ is averaged before or after having been  inverted. The former
procedure leads to a possible ``Hall insulating" behavior.  The latter
 leads to
a $\rho_{xy}$ which grows exponentially when the temperature is lowered.
  Using
the percolating path picture of \cite{ambegaokar}, one  might  speculate
 that the latter
is the proper averaging procedure for the dc limit.

\acknowledgements
The research
was partially supported by the fund for basic research administered by the
Israel
Academy of Sciences and Humanities, the German-Israel Fund
 and by the United States-Israel binational
science foundation (BSF). O. E. W. is grateful to the Einstein Center for
Theoretical Physics of the Weizmann
Institute of Science for partial support.

\end{document}